Tech Science Press

# Blockchain-Enabled EHR Framework for Internet of Medical Things

**Lewis Nkenyereye[1, *], S. M. Riazul Islam[2], Mahmud Hossain[3], M. Abdullah-Al-Wadud[4] and Atif Alamri[4]**

[1]Department of Computer and Information Security, Sejong University, Seoul, 05006, Korea
[2]Department of Computer Science and Engineering, Sejong University, Seoul 05006, Korea
[3]Department of Computer Science, University of Alabama at Birmingham, Alabama, 35294, USA
[4]Department of Software Engineering, King Saud University, Riyadh 11543, Saudi Arabia
[*]Corresponding Author: Lewis Nkenyereye. Email: nkenyele@sejong.ac.kr
Received: XX Month 202X; Accepted: XX Month 202X

**Abstract:** The Internet of Medical Things (IoMT) offers an infrastructure made of smart medical equipment and software applications for health services. Through the internet, the IoMT is capable of providing remote medical diagnosis and timely health services. The patients can use their smart devices to create, store and share their electronic health records (EHR) with a variety of medical personnel including medical doctors and nurses. However, unless the underlying commination within IoMT is secured, malicious users can intercept, modify and even delete the sensitive EHR data of patients. Patients also lose full control of their EHR since most health services within IoMT are constructed under a centralized platform outsourced in the cloud. Therefore, it is appealing to design a decentralized, auditable and secure EHR system that guarantees absolute access control for the patients while ensuring privacy and security. Using the features of blockchain including decentralization, auditability and immutability, we propose a secure EHR framework which is mainly maintained by the medical centers. In this framework, the patients' EHR data are encrypted and stored in the servers of medical institutions while the corresponding hash values are kept on the blockchain. We make use of security primitives to offer authentication, integrity and confidentiality of EHR data while access control and immutability is guaranteed by the blockchain technology. The security analysis and performance evaluation of the proposed framework confirms its efficiency.

## 1 Introduction

The Internet of Medical Things (IoMT) is an extension of Internet of Things (IoT) which combines medical equipment, medical services and medical applications to offer efficient and secure transmission of electronic health records (EHR) within a healthcare system [1]. The EHR data, which are exchanged through the smart devices, can be very beneficial to medical personnel located in remote locations. The IoMT is also projected to offer healthcare services in real time using the internet along with smart devices such as smart phones, smart wearable medical devices or implanted medical devices. Once embraced, the IoMT is expected to reduce the financial cost related to healthcare services while providing timely medical services, quality of medical services, quality of medical experience through fast medical-based decision making [2].

The number of IoT-enabled devices is expected to reach between 50 to 100 billion in the next decade through a variety of internet-based services to enable smart cities, smart homes, smart healthcare, smart grid and so on [3]. Among the different applications of IoT, IoMT is very important as it involves directly the well-being of human kind [4]. With the exponential growth of devices used in IoT, especially in IoMT, user's privacy and security require efficient countermeasures. There are number of security breaches that might occur in IoMT including unauthorized access of patients' EHRs, modification of EHR data, manipulation of EHR data through hijacking of smart medical devices [5,6].

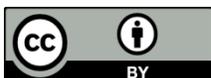

Assume that a patient $A$ is transferred or recommended from hospital $H_1$ to hospital $H_2$. In this scenario, the doctor(s) from $H_2$ may ask the patient to retake the medical tests which were already done in $H_1$ due to untrustworthiness among the medical institutions or probable misinterpretation of medical tests. Retaking the medical test will incur financial burden to the patient while the probability of obtaining the same examination results from the previous hospital is high. Thus, it is appealing to construct an EHR system that allows the health institutions to share the patients data through EHR immutability while the patient keeps full access control of his/her EHR data.

To achieve efficient and secure sharing of EHRs data, there is a number of researchers that suggested security primitives. These techniques mainly focus on constructing an index value for each EHR record and proceed with the encryption of the health data before outsourcing the records through cloud computing. These techniques suffers from the fact that different health institutions use their own techniques of creating the EHR data index or encrypted keys which hinders the sharing among the health institutions [7]. In addition, these systems are based on centralized architectures requiring fully trusted cloud frameworks.

Blockchain technology offers features such as decentralization, immutability, and auditability which can be applied to enforce the privacy and security of EHR in IoMT. Here each medical institution acts as a blockchain node that stores and verifies the health records of patients through the blockchain. Thus, consensus mechanism within the system is used for the audit of the medical treatment process [8].There is also a number of schemes that applied blockchain in EHRs system with the purpose of providing efficient data sharing of health records beside the data confidentiality, integrity and access control [9]. However, these systems suffer from weak privacy preservation which a potential attacker can use to obtain sensitive information such as the preference of certain medical doctor by a number of users due to the public transaction information. Thus, affecting a level of trust among the medical institutions.

In this paper, we propose a blockchain-enabled framework for the internet of medical things. In the proposed scheme, the health institutions act as blockchain nodes, and the blockchain platform is used to store the hash values of the medical records. We also enforce the security of the proposed protocol through security primitives to achieve privacy preservation, authentication and rigorous access control. Specifically, the contribution of the proposed framework can be summarized as follows:

We first design a system model for blockchain-enabled system for the internet of medical things. The proposed architecture allows the different institutions to efficiently and securely share the patients' EHRs using the blockchain technology.

We then design a secure, decentralized, yet auditable framework using Ethereum as a decentralized application platform to offer access control and immutability while the authentication and confidentiality of EHR is achieved using elliptic curve cryptography-based primitives.

We later provide the analysis of the proposed framework by proving the robustness of the framework against well-known malicious attacks. We also provide the performance of the proposed framework through simulation.

The rest of this paper is organized as follows: We first present the related work on current application that are based on blockchain and specifically within the healthcare domain in section 2. In section 3, we describe the system model for the proposed framework and key security primitives used to design the protocol. We describe the different algorithms of the proposed framework in section 4. In section 5, we provide the performance analysis of the framework in terms of security goals and simulation while the concluding remarks and future work are presented in section 6.

**2 Related Work**

Recently, there are several articles investigating the applicability of blockchain technology in areas beside the original financial sector. The authors in [10] provided a survey summarizing the potential fields of application for the blockchain technology. These fields include the smart grid, the healthcare system, the e-voting systems, the insurance field and the education sector. In the same work, the authors discussed the opportunities, the advantages and challenges faced while applying blockchain in the afore-mentioned fields and potential future research directions. In [11], the authors emphasized on a number of technical issues which may affect the privacy, scalability, selfish mining of blockchain-based systems. The work described the current state of the art of blockchain-based systems and future directions. The authors in [12] researched on how the use of blockchain technologies can be maximized in smart cities to form what they labelled as smart communities or smart nations by locating different IoT devices in several geographical sites. The authors also discussed efficient measures for block creation, verification, cryptographic primitives and consensus algorithms.

To attain efficient and secure sharing of EHRs data, a considerable number of articles recommended the adoption of diversified security primitives without the use of blockchain [13-15]. Most of the proposed techniques mainly suggested building an index value for each EHRs records and continue with the encryption of the health data before outsourcing the records through cloud computing. However, different health institutions would use their own techniques of creating the EHRs data index or encrypted keys which would hinder EHRs sharing among the health institutions. Furthermore, Most of these systems were built on a centralized architecture which requires a fully trust of the cloud. There are also a number of work that discussed the integration of blockchain based techniques within the healthcare systems. The authors in [16] discussed the integration of blockchain technology in biomedical and healthcare sector. The authors focused on the advantages that are offered through blockchain including decentralization, security and EHRs data auditability. However, the authors pointed out a number of limitations that would occur by using the current blockchain solutions including EHRs confidentiality, low speed, scalability and potential malicious attacks. Another work on blockchain enabled EHRs systems was proposed in [17]. The authors suggested a scalable system for clinical records through an architecture that was complying with the local health information center. Their studies highlighted a number of drawbacks that might occur including privacy of patient, scalability of EHRs systems mainly due to the volume of EHRs. The authors also discussed the lack of universal standards to enforce the exchange of EHRs on blockchain-based system. There is also a number of schemes applying blockchain in EHRs system with the purpose of providing efficient data sharing of health records beside the data confidentiality, integrity and access control. However, these systems suffer from weak privacy preservation which a potential attacker can use to obtain sensitive data such as the preference of certain medical doctor by a number of users due to the public transaction information. Thus, affecting a level of trust among the medical institutions.

## 3 System Model

In this section, we describe the proposed system architecture along with the main entities in the system as depicted in Fig. 1. The proposed architecture is made by the following main entities. The health authority, the hospital, and the patient. The roles and duties of each entity is described as follows:

The health authority (HA): This is a government office which acts as a trusted authority responsible for regulating all the medical centers within a specific region. Its duties include registering the medical centers based on clear and rigorous rules set by the government. The patient is then registered by the health authority. The health authority will supervise both the patients and the hospitals before joining the blockchain. The health authority is assumed to have severs with sufficient computing capabilities. In case of dispute, the HA is able to trace and reveal the identity of the entities in dispute.

The hospitals: These are public and private health centers such as clinics, community health center, or any other center which may generate and/or make use of the EHRs patients. These hospitals usually have several levels of workers, but for simplicity we only assume two types, namely the doctor and the nurse. These employees perform diversified treatments for a patient including consultation, medical tests, etc. Furthermore, they can create EHRs for patients or access available EHRs for patients.

The patients: These are individuals, a single person or a group of family members, who have access to the EHR generated at a particular medical center during a treatment. The patient has the absolute right on their EHR and can authorize the doctors, nurses or family members to access or update their EHR data.

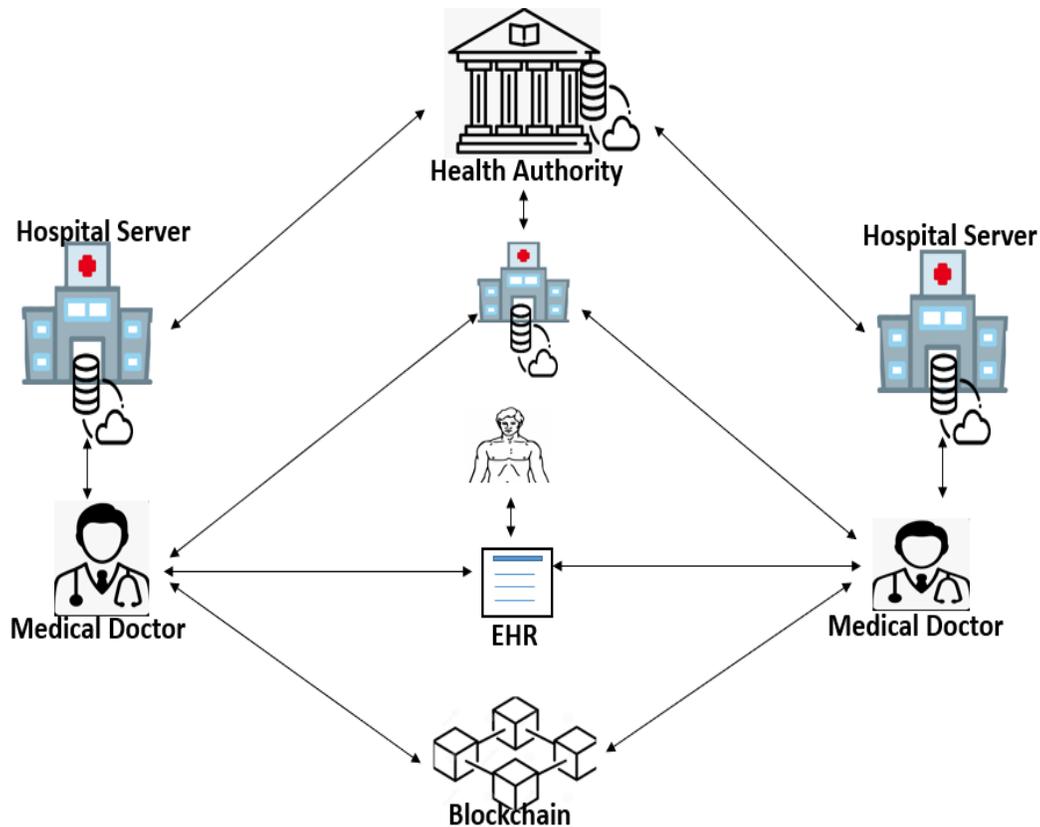

**Figure 1:** Proposed System Architecture

### 3.1 Security Goals

In this subsection, we describe the main security objectives which the proposed protocol aims to satisfy beyond the basic security requirements such as authentication, confidentiality and integrity:

*Privacy Preservation*: The privacy of the patient is very sensitive, thus the proposed protocol should not reveal the real identity of the entities involved in the transactions through it is a public blockchain.

*Unlinkability*: The proposed framework should not allow any unauthorized person or entity to reveal whether two different transactions were originated from the same entity.

*Access control*: The patient should be able to share his/her EHR with the person of his choice using a well-established access control mechanism. Though the EHRs are located in remote servers, none should be able to access the data without a prior authorization of the patient.

### 3.2 Preliminaries

In this subsection, we present the basic security primitives which are used to construct the proposed scheme including blockchain and elliptic curve cryptography.

Blockchain: In this work, we adopt Ethereum as a decentralized platform for smart contracts [18]. We choose Ethereum because (a) it is a programmable blockchain, (2) the users of Ethereum can easily build, deploy and run smart contracts on it and (3) within a transaction, the user can transfer the signed data packet from one account to another. The transactions in Ethereum are made of an account nonce, Gas price,

Gas limit, an address of a recipient amount of ether, the sender digital signature with additional optional fields that a participant can use.

Elliptic Curve Cryptography: Assume an elliptic curve EC and a prime finite field $F_p$. The curve EC can be represented over $F_p$ using a cubic equation $y^2 = x^3 + ax + b$, where $a, b \in F_p$, along with the discriminant $\Delta = 4a^3 + 27b^2 \neq 0$. $P = (x, y)$ is the set of all points over the Curve EC and an additional point $O$ which makes an additive cyclic group $G$ using the point addition + under a tangent method. Assume P and Q to be representing two randoms within the group G. Let $l$ be a line that contains the randoms P and Q, we have a tangent line to the curve EC if $P = Q$ and another point R at the intersection between $l$ and the elliptic curve EC. Let $l'$ be another line that connects R and that line being parallel to the y- axis. We define $P + Q = R' \in E_c$ when we have the line $l'$ intersecting with the curve EC to get a point $R'$. We also define a scalar multiplication as $tP = (P + P + .. + P)$. The details for the complexity assumptions for the Discrete Log (DL) problem, the computational Diffie-Hellman (CDH) problem can be found here for interested readers [19].

## 4 Protocol Description

In this section, we describe the five main stages within the proposed framework. The first phase shows the different steps involved in system initialization, the second and the third phases describes how the entities within our system are registered. The fourth step presents an algorithm that demonstrates how the EHR data are created and shared between the patient and the doctor. The last phase demonstrates how the patient EHR data can be shared between a patient and another doctor with rigorous access control and immutability.

### 4.1 System Setup

The proposed protocol is constructed by adopting Ethereum as the blockchain platform. the cryptographic public parameters are generated as shown in Algorithm 1. This phase is undertaken by the health authority (M) who sets up the system.

**Algorithm 1** *System Setup*

1: Select an elliptic curve E over the finite field $GF$ of prime order $q$
2: Choose $l$ as the corresponding prime order of G
3: Select two hash functions $h_1$ and $h_2$ with $h_1: E \to \mathbb{Z}_p^*$ and $h_2: \{0,1\} \to \mathbb{Z}_q^*$
4: Output the public parameters $PK = \{q, p, E, G, l, h_1, h_2\}$

### 4.2 Patient Registration

A patient participating in the system registers an Ethereum account or address. This account contains the health authority which the corresponding hospital (H) belongs to, the ID of the hospital ($H_i, H_j, ..., H_n$) and the patient ($P_1$). Each entity that registers to the blockchain platform is provided with a key pair $Key = \{SK_{U_i}, PK_{U_i}\}$ as described in Algorithm 2.

**Algorithm 2** *Participant Registration()*

1: Each participant registers Ethereum account $ID_{U_i}$ with $ID_{U_i} = \{\mathcal{M}, \mathcal{H}_i, \mathcal{P}_i\}$
2: The system generate key pair for each participant $Key = \{SK_{U_i}, PK_{U_i}\}$
3: Output $Key = \{SK_{U_i}, PK_{U_i}\}$

### 4.3 Hospital Registration

We assume one hospital but other health institutions such as community health center can join the system. The following algorithm shows the details steps involved in the registration of a medical doctor $\mathcal{D}_i$ working for a specific health institution $\mathcal{H}_i$.

**Algorithm 3**: *MedRegistration ()*

1: Each $\mathcal{D}_i$ or $(ID_{\mathcal{H}_i})$ select $a_i \in \mathbb{Z}_p^*$

2: $\mathcal{D}_i$ computes $A_i$ as the corresponding public key

3: $\mathcal{D}_i$ forwards $N = \{ID_{\mathcal{D}_i}||PK_{\mathcal{D}_i}||A_i\}$ to $\mathcal{M}$ securely and saves $a_i$

4: Upon receiving $N = \{ID_{\mathcal{D}_i}||PK_{\mathcal{D}_i}||A_i\}$, $\mathcal{M}$ performs the validity check

5: $\mathcal{M}$ generates a certificate $Cer_{M,\mathcal{D}_i} = Sign_{SK_\mathcal{M}}(T, ID_{\mathcal{D}_i}, PK_{\mathcal{D}_i}, A_i)$, $T$ being the validity period

6: $\mathcal{M}$ sends $Cer_{M,\mathcal{D}_i}$ securely to $\mathcal{D}_i$

### *4.4 Medical Record Sharing*

Assume that a patient $P_i$ goes to a hospital $H_i$ in which the patient consults a designated medical doctor $D_i$. During the consultation period or treatment, the doctor generates the health records which might include the test results, diagnosis notes and others. The detail of how the health records are generated, encrypted and shared on the servers of the hospital are shown in Algorithm 4.

**Algorithm 4**: HR Sharing ()

1: $\mathcal{D}_i$ generates health records $HR$ of patient $\mathcal{P}_i$

2: $\mathcal{D}_i$ encrypts $HR$ as $CHR_{\mathcal{P}_i} = Enc_{K_{\mathcal{P}_i}}(HR_{\mathcal{P}_i})$, with $K_{\mathcal{P}_i}$ being a symmetric key

3: $\mathcal{D}_i$ stores $CHR_{\mathcal{P}_i}$ in the server $\mathcal{C}_1$ belonging to $H_i$

4: $H_i$ generates the corresponding hash value $eh_1 = h_2(CHR_{\mathcal{P}_i})$

5: $\mathcal{D}_i$ generates a transaction record $\mathcal{R}_1 = \{T_1, T_{y1}, eh_1\}$ with $T_{y1}$ being a transaction type

6: $\mathcal{P}_i$ generates $\mathcal{X}_i = h_2(\mathcal{H}_1||T_1||0, k_t), \mathcal{Y}_i = h_2(\mathcal{H}_1||T_1||1, k_t)$, $k_t$ being a random chosen by $\mathcal{P}_i$

7: $\mathcal{P}_i$ also computes $\mathcal{Z}_i = tx_{id} \oplus \mathcal{Y}_i, \mathcal{K}_i = K_{\mathcal{P}_i} \oplus \mathcal{Y}_i$, $tx_{id}$ being the transaction ID of $\mathcal{R}_1$

8: $\mathcal{D}_i$ then sends $\{\mathcal{X}_i, \mathcal{Y}_i, \mathcal{K}_i\}$ to the server $\mathcal{C}_1$

### *4.5 Medical Record Resharing*

Assume that the patient is sharing his health records to another doctor $D_j$ within the same hospital. However, the algorithm can easily be adopted for sharing the health records in a different hospital. The detail of the sharing steps are described in algorithm 5.

**Algorithm 5**: HR Sharing

1: $\mathcal{P}_i$ selects a random $r_\tau \in \mathbb{Z}_p^*$

2: $\mathcal{P}_i$ computes a tag $ST = h_1(r_\tau A_j)G$ and $R_\tau = r_\tau G$, $A_j$ being the public key of $\mathcal{D}_j$

3: $\mathcal{P}_i$ generates an authorization pass $A_{pass} = (\mathcal{H}_1, T_1, k_t)$

4: $\mathcal{P}_i$ computes $C_1 = Enc_{PK_{\mathcal{D}_2}}(A_{pass})$

5: $\mathcal{P}_i$ generate a transaction $Trans = (R_\tau||ST||C_1)$ and broadcast it to the Ethereum

6: $\mathcal{D}_2$ computes $ST' = h_1(a_j R_\tau)G$ and check if $ST' = ST$

7: $\mathcal{D}_2$ computes $v = Dec_{SK_{\mathcal{D}_2}}(C_1)$ to recover $A_{pass}$

8: $\mathcal{D}_2$ computes $\mathcal{X}'_i = h_2(\mathcal{H}_1||T_1||0, k_t)$ and $\mathcal{Y}'_i = h_2(\mathcal{H}_1||T_1||1, k_t)$

9: $\mathcal{D}_2$ computes $\vartheta = Sig_{SK_{\mathcal{D}_2}}(\mathcal{W})$ with $\mathcal{W} = \mathcal{X}'_i$

10: $\mathcal{D}_2$ sends $\psi$ to $\mathcal{H}_1$ with $\mathcal{H}_1 = (\mathcal{W}||\vartheta||Cer_{\mathcal{H}_1,\mathcal{D}_2})$ with $Cer_{\mathcal{H}_1,\mathcal{D}_2}$ being the certificate of $\mathcal{D}_2$

11: Upon receiving the message, $\mathcal{H}_1$ checks the validity of the certificate

12: $\mathcal{H}_1$ proceed with the verification of the signature $\vartheta$
13: $\mathcal{H}_1$ sends $(\mathcal{Z}_i, \mathcal{K}_i, CHR_{\mathcal{P}_i})$ to $\mathcal{D}_2$
14: Upon receiving $(\mathcal{Z}_i, \mathcal{K}_i, CHR_{\mathcal{P}_i})$, $\mathcal{D}_2$ computes $tx'_{id} = \mathcal{Z}_i \oplus \mathcal{Y}'_i$ and $K'_{\mathcal{P}_i} = \mathcal{K}_i \oplus \mathcal{Y}'_i$
15: $\mathcal{D}_2$ recovers $eh'_1$ based on $tx'_{id}$ and v
16: Verifies if $eh'_1 = h_2(CHR_{\mathcal{P}_i})$?
17: $\mathcal{D}_2$ computes $\alpha = Dec_{K'_{\mathcal{P}_i}}(CHR_{\mathcal{P}_i})$ to recover $HR_{\mathcal{P}_i}$

We assume in this work an optional time line after which a given $HR$ can be deleted. The server $\mathcal{H}_1$ can therefore delete the health records. In addition, the corresponding hash values $CHR_{\mathcal{P}_i}$ shall also be deleted from the blockchain.

## 5 Performance

In this section we present the performance of the proposed scheme which is divided in two main parts. The first subsection discusses the security analysis of the protocol and the next subsection provides the evaluation of the scheme through simulation.

### 5.1 Security Analysis

In this subsection, we discuss how the proposed scheme is secure in terms of the security goals.

*Privacy Preservation*: The proposed protocol is secure against the leaking of the sensible IDs of all the entities in the system though all the blockchain transactions are public. As shown in Algorithm 5, we build an authorization token $A_{pass}$ in which the address of the medical doctor was omitted. Thus, even if an adversary gets hold of the contents, he cannot retrace the real entity of the doctor. Thus, the proposed scheme ensures the privacy of the patients and the doctors.

*Authentication*: In the proposed scheme, a patient first books an appointment with a doctor using the registration credentials. As mentioned in Algorithm 1, each participant registers an Ethereum account $ID_{\mathcal{U}_i} = \{\mathcal{M}, \mathcal{H}_i, \mathcal{P}_i\}$ and is provided with key pairs $Key = \{SK_{\mathcal{U}_i}, PK_{\mathcal{U}_i}\}$. Therefore any unregistered patient cannot even successfully register an appointment with the doctor $D_i$ through the hospital $\mathcal{H}_i$. Thus, the proposed framework guarantees the authentication of entities.

*Authorization*: The proposed scheme achieves authorization because the patient $\mathcal{P}_i$ can only share his EHR data with designated doctors. $\mathcal{P}_i$ generates a transactions token $Trans = (R_\tau || ST || C_1)$ within the Ethereum blockchain which is sent to the doctor $\mathcal{D}_2$. $\mathcal{D}_2$ then proceeds with the verification of the authorization token using its private key $SK_{\mathcal{D}_2}$. Based on elliptic curve discrete logarithm problem the (ECDLP), a non-valid secret key cannot achieve a valid verification.

*Confidentiality*: As shown in Algorithm 4, each transaction between the patient $\mathcal{P}_i$ and any doctor $\mathcal{D}_i$ is encrypted. $\mathcal{P}_i$ encrypts the authorization token as $C_1 = Enc_{PK_{\mathcal{D}_2}}(A_{pass})$ and sends to the doctor $\mathcal{D}_1$. Even if a malicious user gets hold on the message, he cannot get any meaningful content from it. Therefore, the proposed framework guarantees the confidentiality of patient's sensitive data.

*Auditability*: The proposed scheme achieves strong auditability which a doctor may not deny in case the patient EHR data are tampered with. It is hard or almost impossible for an adversary to modify the EHR which is exchanged between the patient and the doctor. In algorithm 2, we see that the doctor generates a transaction $\mathcal{R}_1 = \{T_1, T_{y1}, eh_1\}$ and broadcast it to the Ethereum blockchain which guarantees the immutability of the transaction

### 5.2 Performance Analysis

In the performance evaluation, we used Ethereum as a public blockchain, this is mainly because the transactions fee are very affordable compared to other public blockchain. In addition, the time to upload in Ethereum varies between 3 to 4 min, which is quite convenient compared to other public blockchain such Bitcoin. We performed the evaluation of the proposed framework using a Windows 10 Home system, 3.6 GHz with i7 CPU for 16 GB RAM capacity. The size of the Ethereum address for $ID_{U_i}$ is set to 25 bytes while the hash functions for $h_2(.)$ and $h_3(.)$ are SHA-256 and SHA-3 respectively. The key pair $Key = \{SK_{U_i}, PK_{U_i}\}$ are 32 bytes for public key and 66 for the private key. The symmetric keys used in the protocol are set to 128 bits. We consider secp256k1 algorithm for the operations related to encryption/ decryption $Enc_K(.)/Dec\_k(.)$ and signature generation and verification $Sig_{SK}(.)/Ver_{PK}(.)$. We adopt the parameter settings from this work [20] for transactions and certificate calculation.

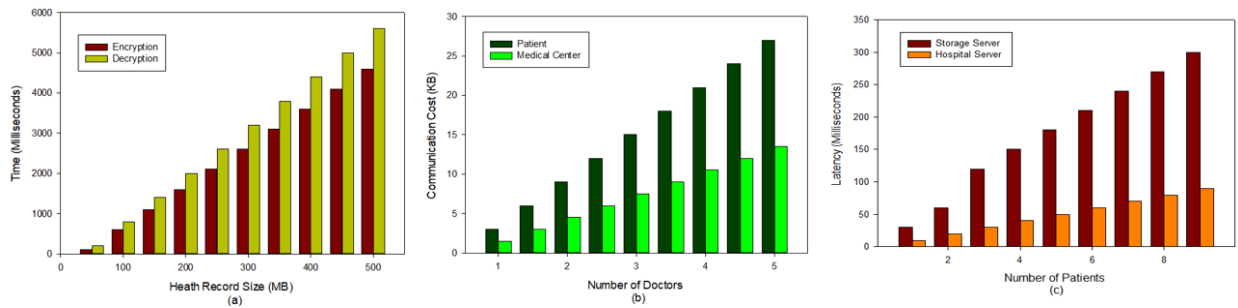

Figure 2: (a) Encryption /Decryption performance based on the size of health records, (b) Communication cost for patient/hospital based on the number of doctors, (c) latency performance for hospital /cloud based on the number of patient

We first investigate the computation cost for the encryption and decryption phases based on the size of the health records. This is quite crucial since some EHR may have multimedia contents such as pictures. We considered the health records with a range of 50 to 500 MB since we assume that the EHR can contain multimedia contents. As seen in Fig. 2a, the computation overhead increases as the size of the EHR increases. Then we investigate the communication cost incurred by the patient device and the hospital devices during the diagnosis phase. As seen on the Fig. 2b, it takes about 4 KB on the device of the patient when he is being diagnosed by a single doctor. On the hospital size, since we assume that the hospital sever has sufficient computation cost, the communication cost incurred on the device is very minimal as depicted in Fig. 2b. Lastly we investigate the computational burden on the hospital servers based on the number of the patients. As shown in Fig. 2c, for a number of around 10 patients making the consultation at the same time, the computation burden is around 300 ms, which is quite satisfactory.

## 6 Conclusion

In this paper, we presented a secure and blockchain enabled framework for EHR data for the internet of medical things (IoMT). The system architecture of the proposed framework allows a patient to secure share his data with several medical personnel with rigorous access control and immutability. We use a consortium blockchain to store the hash values of the patients data to offer auditability and immutability of the EHR data and ensured patient's control on his sensitive data. We made use of cryptographic primitives to guarantee the robustness of the proposed framework against well-known attacks. The security analysis of the proposed framework meets the requirements including privacy, unlinkability and access control. Furthermore, the performance evaluation shows that the proposed framework yields high computational efficiency.

**Funding Statement:** The authors are grateful to the Deanship of Scientific Research at King Saud University for funding this work through Vice Deanship of Scientific Research Chairs: Chair of Pervasive and Mobile Computing.

**Conflicts of Interest:** The authors declare that they have no conflicts of interest to report regarding the present study.

**References**

[1] A. Gatouillat, Y. Badr, Y. Massot and E. Sejdić, "Internet of medical things: a review of recent contributions dealing with cyber-physical systems in medicine," *IEEE Internet of Things Journal*, vol. 5, no. 5, pp. 3810-3822, 2018.

[2] F. Alsubaei, A. Abuhussein, V. Shandilya and S. Shiva, "IoMT-SAF: internet of medical things security assessment framework," *Internet of Things*, vol 8, pp. 100-123, 2019.

[3] S. Li, L.D. Xu and S. Zhao, "5G Internet of Things: a survey," *Journal of Industrial Information Integration*, vol. 10, pp. 1-9, 2018.

[4] F.A. Alaba, M. Othman, I.A. Hashem and F. Alotaibi, "Internet of things security: a survey," *Journal of Network and Computer Applications*, vol. 88, pp. 10-28, 2017.

[5] S. Pirbhulal, O.W. Samuel, W. Wu, A.K. Sangaiah and G. Li, "A joint resource-aware and medical data security framework for wearable healthcare systems," *Future Generation Computer Systems*, vol. 95, pp. 382-391, 2019.

[6] R. Hamza, Z. Yan, K. Muhammad, P. Bellavista and F. Titouna, "A privacy-preserving cryptosystem for IoT E-healthcare," *Information Sciences*, vol. 527, pp. 493-510, 2020.

[7] C. Agbo, C. Mahmoud, H. Qusay and J.M. Eklund, "Blockchain technology in healthcare: a systematic review," *Healthcare*, vol. 7, no. 2, pp 56-71, 2019.

[8] T.T. Kuo, H.E. Kim and L. Ohno-Machado, "Blockchain distributed ledger technologies for biomedical and health care applications," *J Am Med Inform Assoc*, vol. 24, no. 6, pp. 1211–1220, 2017.

[9] S. Cao, G. Zhang, P. Liu, X. Zhang and F. Neri, "Cloud-assisted secure e-health systems for tamper-proofing ehr via blockchain," *Information Sciences*, vol. 485, pp. 427–440, 2019.

[10] A. Monrat, O. Schelen and K. Andersson, ''A survey of blockchain from the perspectives of applications, challenges, and opportunities,'' *IEEE Access*, vol. 7, pp. 117134–117151, 2019.

[11] Z. Zheng, S. Xie, H. N. Dai, X. Chen and H. Wang, ''Blockchain challenges and opportunities: a survey,'' *Int. J. Web Grid Services*, vol. 14, no. 4, pp. 352-364, 2018.

[12] S. Aggarwal, R. Chaudhary, G. S. Aujla, N. Kumar, K.K.R. Choo and A. Y. Zomaya, ''Blockchain for smart communities: applications, challenges and opportunities,'' *J. Netw. Comput. Appl.*, vol. 144, pp. 13–48, 2019.

[13] Z. Yuan, C. Xu, H. Li, Y. Kan and X. Lin, "Healthdep: an efficient and secure deduplication scheme for cloud-assisted ehealth systems," *IEEE Transactions on Industrial Informatics*, vol. 14, no. 9, pp. 4101–4112, 2018.

[14] J. Sun and Y. Fang, "Cross-domain data sharing in distributed electronic health record systems," *IEEE Transactions on Parallel & Distributed Systems*, vol. 21, no. 6, pp. 754–764, 2010.

[15] Y.S. Rao, "A secure and efficient ciphertext-policy attribute-based signcryption for personal health records sharing in cloud computing," *Future Generation Computer Systems*, vol. 67, pp.133-151, 2017.

[16] M.S. Sahoo and P. K. Baruah, ''HBasechainDB: a scalable blockchain framework on Hadoop ecosystem,'' *Supercomputing Frontiers*, pp. 18–29, 2018.

[17] P. Zhang, J. White, D. C. Schmidt, G. Lenz and S. T. Rosenbloom,''FHIRChain: applying blockchain to securely and scalably share clinical data,'' *Comput. Struct. Biotechnol. J.*, vol. 16, pp. 267–278, 2018.

[18] K.N. Griggs, O. Ossipova, C.P. Kohlios, A.N. Baccarini, E.A. Howson *et al.*, "Healthcare blockchain system using smart contracts for secure automated remote patient monitoring," *Journal of medical systems*, vol.42, no. 7, p.130 -144, 2018.

[19] K. Mahmood, S.A. Chaudhry, H. Naqvi, S. Kumari, X. Li *et al.*, "An elliptic curve cryptography based lightweight authentication scheme for smart grid communication," *Future Generation Computer Systems*, vol. 81, pp.557-565, 2018.

[20] A. Celesti, A. Ruggeri, M. Fazio, A. Galletta, M. Villari *et al.*, "Blockchain-based healthcare workflow for tele-medical laboratory in federated hospital IoT clouds," *Sensors*, vol. 20, no. 9, p.2590-2599,2020.